\documentclass[submission,copyright,creativecommons]{eptcs}
\providecommand{\event}{TERMGRAPH 2022}
\usepackage{underscore}

\usepackage{listings}
\usepackage{alltt}
\usepackage{xcolor}
\usepackage{lmodern}
\usepackage[T1]{fontenc}
\usepackage{amsmath}
\usepackage{amsthm}
\usepackage{amssymb}
\usepackage{stmaryrd}
\usepackage{mathtools}
\usepackage{cmll}
\usepackage{mathpartir}
\usepackage{enumitem}
\usepackage{comment}
\usepackage{wrapfig}
\usepackage{subfig}
\usepackage{textgreek}
\usepackage{footmisc}
\usepackage[scaled=0.85]{roboto-mono}
\usepackage{makecell}
\usepackage{enumitem}
\usepackage{booktabs}
\usepackage{multirow}
\usepackage{multicol}
\usepackage{graphicx}

\theoremstyle{definition}
\newtheorem{definition}{Definition}

\DeclarePairedDelimiter\trans{\llparenthesis}{\rrparenthesis}

\newcommand{\roo}[0]{\leftrightarrow}

\newcommand{\rmo}[0]{\to}
\newcommand{\rmoz}[0]{\rightharpoonup}
\newcommand{\rmm}[0]{\asymp}
\newcommand{\cartprod}[0]{\otimes}
\newcommand{\bigcartprod}[0]{\bigotimes}

\newcommand{\thelangname}{Ideograph}
\newcommand{\thetitle}{\thelangname{}: A Language for Expressing and Manipulating Structured Data}
\newcommand{\thetitlesplit}{\thelangname{}: A Language for Expressing and Manipulating Structured Data}

\newcommand{\thetermname}{ideogram}

\newcommand{\relatedworkspacefudge}{\vspace{-0.11em}}

\title{\thetitlesplit}
\author{Stephen Mell
\institute{University of Pennsylvania}
\email{sm1@cis.upenn.edu}
\and
Osbert Bastani
\institute{University of Pennsylvania}
\email{obastani@seas.upenn.edu}
\and
Steve Zdancewic
\institute{University of Pennsylvania}
\email{stevez@seas.upenn.edu}}

\def\titlerunning{\thetitle{}}
\def\authorrunning{S. Mell, O. Bastani \& S. Zdancewic}

\begin{document}
\maketitle

\begin{abstract}
We introduce \thelangname{}, a language for expressing and manipulating structured data.
Its types describe kinds of structures, such as natural numbers, lists, multisets, binary trees, syntax trees with variable binding, directed multigraphs, and relational databases. Fully normalized terms of a type correspond exactly to members of the structure, analogous to a Church-encoding. Moreover, definable operations over these structures are guaranteed to respect the structures' equivalences. In this paper, we give the syntax and semantics of the non-polymorphic subset of \thelangname{}, and we demonstrate how it can represent and manipulate several interesting structures.
\end{abstract}

\section{Introduction}
\label{sec:intro}
Structured data is ubiquitous: lists, trees, graphs, relational databases, and syntax trees are just a few of the structures that underpin computer science. We often want to perform operations on such objects in ways that both respect and leverage their structure. For instance, we might wish to aggregate the elements of a bag (multiset). We could represent bags as lists and fold over them as lists, but this provides no guarantee that the result is invariant to the order.
Or, we might wish to manipulate syntax trees of programs. We could represent variables as names or de Brujin indices~\cite{debruijn}, but in either case operations on the representation must be shown to respect the binding structure. Other, similar, circumstance arise often in practice.

Yet, there are surprisingly few formalisms for actually defining such structures, much less for defining invariant-respecting operations over them. Relational database schemas define bags of records, with certain additional structure (most notably, foreign key constraints between tables). Most widely used programming languages, like C, Java, and Python, and data interchange formats, like Google’s Protocol Buffers, have limited type systems, supporting at most product, sum, and function types, but not supporting the graph structures and constraints that would be required to define bags or syntax trees with variable binding. Dependently-typed languages, like Coq, are capable of imposing complex constraints, but even simple data structures, like syntax trees with binding structure, have proven tricky to deal with in practice~\cite{poplmark}.

As a result, we resort to implementing ad-hoc solutions. For aggregating over bags, we can separately prove that our function is invariant to order, and thus is truly a function over bags rather than lists. For manipulating syntax trees, we can separately prove that our substitution operation is capture-avoiding. However, this must be done for each new structure and operation. We want a general formalism for representing and manipulating a rich class of structures.

Graphs are a common formalism for encoding many kinds of data, but they don't capture everything. For example, binary trees are ``graphs'', but they have more structure: they have two distinct kinds of nodes (``branch'' and ``leaf''), two kinds of edges (``left child'' and ``right child''), and the requirements that (1) each branch has one left child and one right child, and (2) that every node except the root has one parent. The formalism of graphs also does not account for manipulations: given a binary tree whose leaves are themselves tagged with other binary trees, we might want to collapse this tree of trees into a single tree. While a good starting point, graphs \emph{per se} are not a precise enough formalism to capture these structures and operations.

Church-encodings~\cite{tapl,encodetermalgebras} in polymorphic lambda calculus can precisely express many such structures, and they provide a natural notion of structure-respecting manipulation. For example, the type $\forall X.\ (X \to X \to X) \to (Y \to X) \to X$ encodes exactly binary trees whose leaves are labeled with elements of $Y$. (Roughly, the two arguments correspond to the two kinds of nodes in binary trees: $X \to X \to X$ corresponds to branch nodes, with two tree-children and one parent; $Y \to X$ corresponds to leaf nodes with one $Y$-child and one parent.) Further, Church-encodings of structures are themselves functions, corresponding to generalized fold operations: to use a term, you provide one function per constructor, and each occurrence of a constructor in the term is replaced by the corresponding function call. This allows the manipulation of terms in a structure-respecting way.
However, standard Church-encodings~\cite{encodetermalgebras} are over heterogeneous term algebras, but bags, relational database schemas, and syntax trees with variable binding are not term algebras.
Finally, these encodings are not canonical, e.g., $\forall X.\ (Y \to X) \to (X \to X \to X) \to X$ also encodes binary trees. As the complexity of encoded structures increases, the number of equivalent encodings may increase combinatorially.

In this work, we leverage the complementary strengths of these two approaches, building a language called \thelangname{}, where both the terms and the types are graph-structured. By having a calculus, we are able to precisely encode many structures and define structure-respecting operations over them. By having terms that are graphs rather than trees, we are able to capture a richer set of structures. By having types that are graphs, we are able to eliminate many redundant encodings of structures.

We begin by using examples to describe the terms (Section~\ref{sec:terms}), operational semantics (Section~\ref{sec:semantics}), types (Section~\ref{sec:types}), and a well-formedness condition (Section~\ref{sec:validity}), followed by the formalism (Section~\ref{sec:formalism}). We then present representations of several data structures in the language (Section~\ref{sec:repr}) and demonstrate the manipulation of such structures (Section~\ref{sec:manip}). We conclude with discussions of related work (Section~\ref{sec:relatedwork}) and future work (Section~\ref{sec:futurework}). For clarity and concision, we omit polymorphism from this presentation.

\section{\thelangname{}}
\label{sec:syntax:zero}
\begin{figure}[!t]
    \footnotesize
    \centering
    \centerline{\includegraphics[width=6.3in]{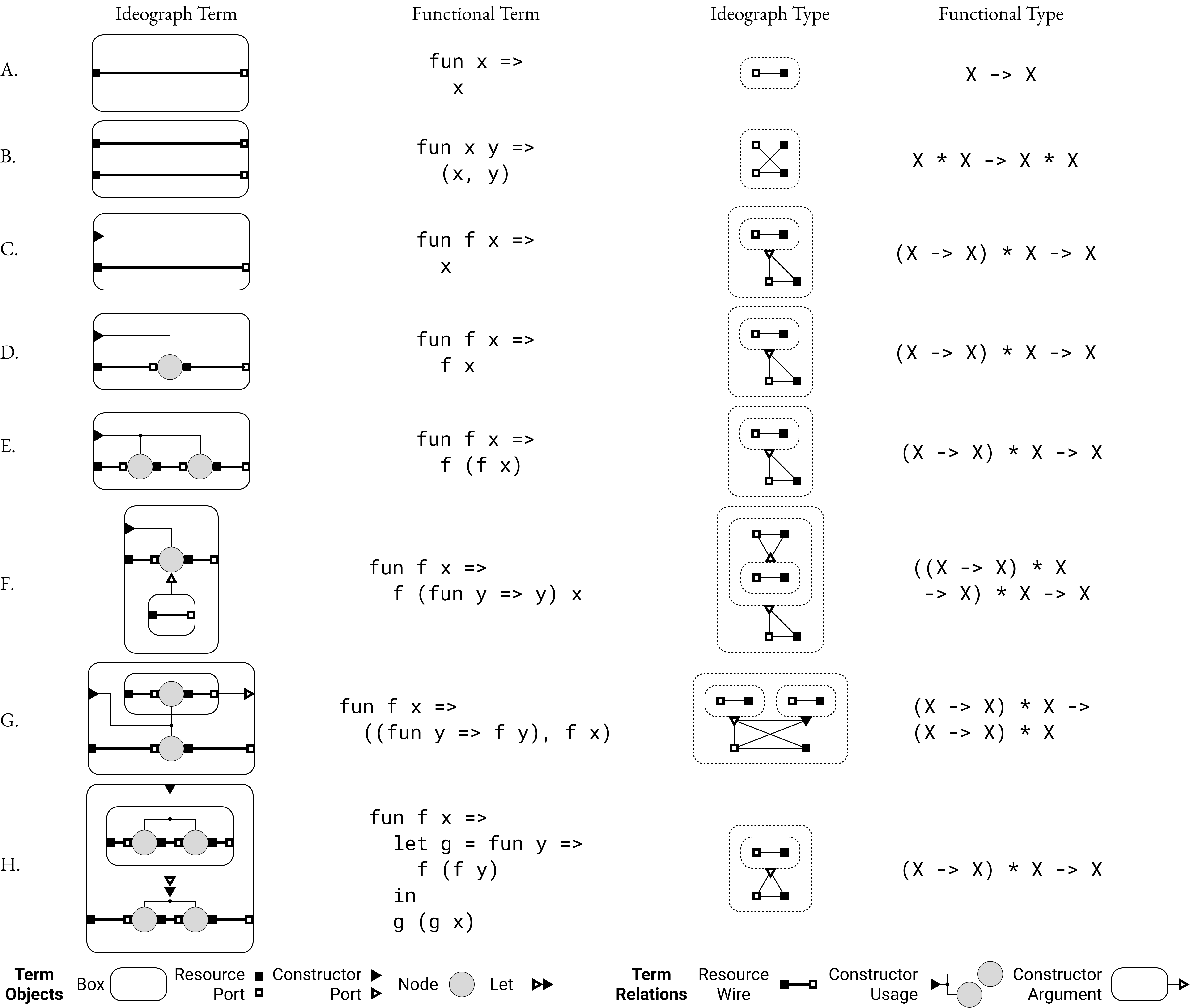}}
    \caption{\small
Terms in \thelangname{}, along with their types, and analogues of each in a generic functional programming language. Though there are multiple Ideograph types that could represent the same functional type, these should provide the right intuition. More precisely, \thelangname{}'s type system is linear in the sense of Girard~\cite{linearlogic}, and here we translated standard function types $X \to Y$ as $!{(X \multimap Y)}$. This makes function-typed arguments reusable, while other arguments are linear. However, there are other translations~\cite{twotrans}, and this was a purely expository choice. For simplicity of presentation, we only use one primitive (\texttt{X}) in the types, so the types given are not necessarily the principal types of their terms. The formalism allows labeling resource fields with additional primitives.}
    \label{fig:translation-table}
\end{figure}

\thelangname{} is a means of expressing and composing structured graphs. Its terms are ``structured'' in the sense of having distinct kinds of edges and nodes. Nodes have ``ports'', and edges connect nodes via these ports. Though we introduce additional constructs to support computation and polymorphism, the core idea is to substitute copies of a graph for certain nodes in another graph. Because the formal definitions of the syntax and semantics have many moving parts and are opaque without context, we begin by stepping through the examples in Figure \ref{fig:translation-table}, which demonstrate the key aspects of \thelangname{}. The formalism is presented in Section~\ref{sec:formalism}.

\subsection{Terms}
\label{sec:terms}
A term in \thelangname{}, henceforth an \emph{\thetermname{}}, consists of a set of \emph{boxes} ($\mathcal{B}$, depicted as rounded rectangles e.g. in Figure \ref{fig:translation-table}), \emph{nodes} ($\mathcal{N}$, gray circles or rectangles), and \emph{ports} ($\mathcal{P}$, small triangles and squares, hollow or filled), with several relations among these different objects. The boxes, nodes, and ports reside in other boxes (the relation $R_{R}$, depicted by nesting), with the boxes forming a tree. Each port is either a \emph{receiver} ($\mathcal{P}_-$, hollow) or a \emph{provider} ($\mathcal{P}_+$, filled) and is for either a \emph{resource} ($\mathcal{P}_R$, square) or a \emph{constructor} ($\mathcal{P}_C$, triangle).
Ports are typically \emph{attached} ($R_A$, depicted by contact) to a node or a box. We will introduce other relations between these objects as they arise in the following examples. Figure~\ref{fig:formalexg} and Figure~\ref{fig:formalopsemformal} give illustrations annotated with these objects and relations. Terms are also subject to a well-formedness condition that will be discussed in Section \ref{sec:validity}.
The comprehensive formalism is deferred to Section~\ref{sec:formalism}.

\paragraph{Simple functions.}
Consider the identity function (Example A in Figure \ref{fig:translation-table}). As an \thetermname{}, it is a box with two resource ports: one resource provider port (solid square), analogous to the input \texttt{x} of the functional analogue; and one resource receiver port (hollow square), analogous to the function output. Because the identity function returns its input as its output, there is a wire between the two resource ports, which is captured by the bijective \emph{resource wiring relation} ($R_{WR}$, depicted with thick lines) between resource provider and receiver ports.

Example B is slightly more complicated. It has two resource provider ports for the two arguments, \texttt{x} and \texttt{y}, and two resource receiver ports for the two outputs, the left and right sides of the tuple. The two wires indicate which input gets returned as which output.
Because the resource wiring relation is bijective, there are no terms of this type analogous to returning the pairs \texttt{(x, x)} or \texttt{(y, y)}. This makes resources linear.

\paragraph{Calling functions.}
Example C is the same as Example A, but with the addition of a single, unused constructor provider port (depicted as a filled triangle, sometimes just called a ``constructor''), corresponding to the unused argument \texttt{f} with type \texttt{X -> X}. This lack of use is allowed because constructors are not linear in the way that resources are.

Examples D and E are more interesting, as they actually use the constructor. Nodes are analogous to function invocations, and so in Example D, we have one node corresponding to the one call to \texttt{f}. To capture that the node was constructed by the constructor provider port, the port and node are associated by the \emph{constructor usage relation} ($R_{CU}$, depicted with a thin, possibly branching line; the branching structure is not meaningful, and exists to improve readability).
Each node must be associated with exactly one constructor, but constructors can be associated with any number of nodes. In Example E, the constructor is used to construct two nodes, analogous to the two calls to \texttt{f}.

Nodes can have associated ports in the same way that boxes can. In Examples D and E, the nodes each have one resource receiver port, corresponding to the input to the \texttt{f} call, and one resource provider port, analogous to the output. In Example D, the wire on the left corresponds to passing the input, \texttt{x}, to the call to \texttt{f}, and the wire on the right is analogous to returning the output of \texttt{f}. In Example E, the wires correspond to passing \texttt{x} to the first call to \texttt{f}, passing its output to the second call to \texttt{f}, and finally returning its output. Note that nodes and boxes have opposite views of provider and receiver: when calling a function (analogous to a node), the function receives the input and provides the output; when defining a function (analogous to a box), the context provides the input and receives the output.

\paragraph{Passing and returning functions.}
Example F is like Example D, but instead of \texttt{f} taking a single argument of type \texttt{X}, it also takes an argument of type \texttt{X -> X}. This is analogous to the constructor receiver port (hollow triangle) attached to the node. That we are passing the identity function corresponds to the nested box, a copy of Example A, that is connected to the constructor receiver port by the \emph{constructor argument relation} ($R_{CA}$). This relation is a bijection between constructor receiver ports and boxes (excluding the top-level box). This is the first example with non-trivial box-residence structure: there are two boxes, one (depicted as inner) being the child of the other (depicted as outer).

Example G shows how a function can return a function: the function-typed output is analogous to the constructor receiver port, which, as in Example F, must be connected to a box.
Note that the constructor usage relation can sometimes cross box boundaries, analogous to lexical scoping for functions: in Example G, the constructor provider port corresponding to \texttt{f} is used in the inner box. Formally, the constructor usage relation can associate nodes to constructor provider ports in the same box or one that is higher in the residence tree.

\paragraph{Let-bindings.}
So far we we have only seen values. To have terms that can take operational steps, we also have let-bindings ($\mathcal{D}$, depicted by the contact of a constructor receiver port and a constructor provider port). In Example H, the box connected to the receiver port is analogous to the body of the let-binding, \texttt{fun y => f (f y)}, and the two nodes connected to the provider port are analogous to the instances of \texttt{g}. Each let-binding must be attached to exactly one constructor receiver port and one constructor provider port. Each port must be associated with exactly one box, node, or let-binding. Each let-binding also has a type, discussed in Section~\ref{sec:types}. 

\subsection{Operational Semantics}
\label{sec:semantics}
Recall that the core idea of \thelangname{} is to substitute terms for the nodes of other terms. The let-binding construct connects a box (the binding's body) to some nodes (the occurrences of the binding's bound variable). The single reduction rule of \thelangname{} is the substitution of the body for the occurrence nodes. (With polymorphism, there is also a type-level let-binding, and there is a second reduction rule for substituting at the type level.)

\begin{figure}[b]
    \footnotesize
    \centering
    \centerline{\includegraphics[width=6.3in]{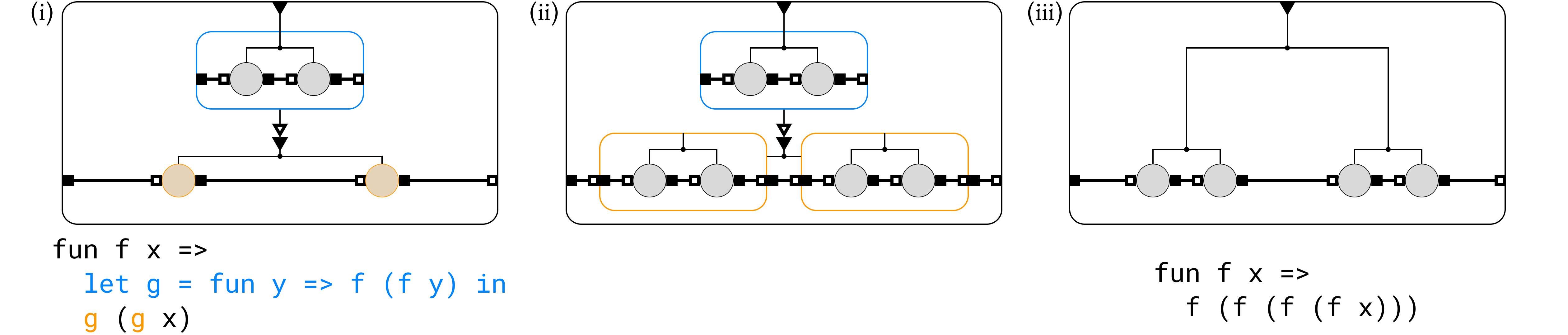}}
    \caption{\small (i) and (iii) depict \thetermname{}s and their functional analogues. (i) evaluates to (iii) in one step, by inlining the let-binding. (ii) is not a term, but depicts how inlining is done. Colors indicate analogies.}
    \label{fig:operational-semantics-simple}
\end{figure}

Consider Figure \ref{fig:operational-semantics-simple}~(i). The let-binding is analogous to the definition of \texttt{g}, sequencing two nodes, each analogous to a call to \texttt{f}. This let-binding is then used to construct two nodes which are themselves sequenced. Stepping takes the contents of the box (blue) and places a copy of it at each occurrence node (orange). This intuition is depicted in Figure \ref{fig:operational-semantics-simple}~(ii), but requires a bit of clean-up. Each adjacent pair of resource receiver and provider ports is replaced with a wire. The \texttt{f} nodes in the body of \texttt{g} remain as \texttt{f} nodes even after substitution, though we now have four of them. Finally, we erase the let-binding, to get Figure \ref{fig:operational-semantics-simple}~(iii).

\begin{figure}[!t]
    \footnotesize
    \centering
    \centerline{\includegraphics[width=6.3in]{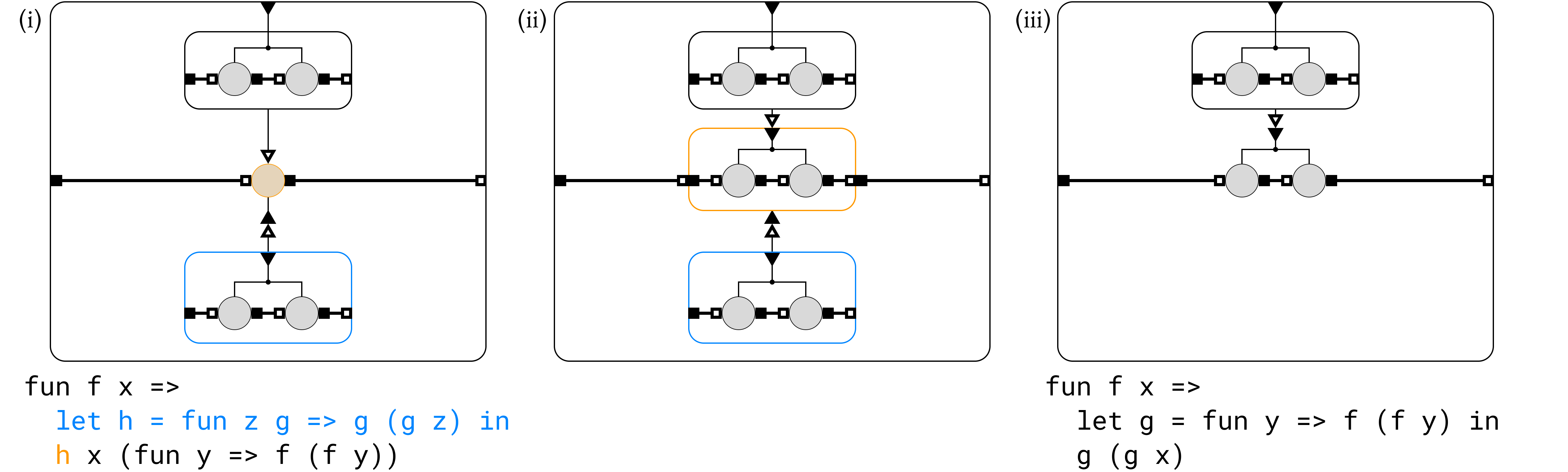}}
    \caption{\small (i) and (iii) depict \thetermname{}s and their functional analogues. (i) evaluates to (iii) in one step, by inlining the let-binding. (ii) is not a term, but depicts how inlining is done. Colors indicate analogies.
    Figure~\ref{fig:formalopsemformal} contains a formally annotated version of this example.}
    \label{fig:operational-semantics-higher}
\end{figure}

Though capturing the core idea of substituting terms for nodes, the previous example does not have constructor ports on the let-binding. Figure~\ref{fig:operational-semantics-higher} does. We again copy the body of the let-binding, \texttt{h}, for its occurrence nodes, and then we erase the let-binding and replace each pair of resource receiver and provider ports with a wire. Crucially, the pair of constructor receiver and provider ports becomes a new let-binding. This means that the term can continue stepping. (Indeed, Figure~\ref{fig:operational-semantics-higher}~(iii) is the same term as Figure~\ref{fig:operational-semantics-simple}~(i).) For a formal definition of the operational semantics and a formalization of this example, see Section~\ref{sec:formalism}.

\subsection{Types and Correspondences}
\label{sec:types}
Doing substitution as outlined above poses a challenge: how do we know the correspondence between the ports on the box and the ports on a node? Figure~\ref{fig:operational-semantics-simple} assumes that the left ports and right ports on the nodes correspond to the left port and right port on the box, respectively. The primary role of types in \thelangname{} is to make this correspondence precise.

We now define a \emph{type} and, between components of a type ($\mathcal{I}$, $\mathcal{F}$, defined shortly) and components of a term ($\mathcal{B}$, $\mathcal{N}$, $\mathcal{P}$, $\mathcal{D}$), a \emph{correspondence relation}. The counterparts of the ports in a term are the \emph{fields} ($\mathcal{F}$, depicted the same as ports) of a type, and a correspondence relation maps each port to at most one field. In a term, ports are attached to a box or a node, whereas in a type, fields reside in an \emph{interface} ($\mathcal{I}$, depicted as dotted, rounded rectangles), and a correspondence relation maps each box and each node to at most one interface. The residence of fields in an interface, as well as the nesting of interfaces, is captured by the \emph{residence relation} ($R_R$, depicted by containment), much like it is for terms. Correspondences must be consistent, in that if a box or node corresponds to an interface, then the ports attached to the box or node must correspond bijectively to the fields in the interface. Like ports, fields are either received ($\mathcal{F}_{-}$) or provided ($\mathcal{F}_{+}$) and are for either constructors ($\mathcal{F}_{C}$) or resources ($\mathcal{F}_R$).
Constructor and resource ports correspond to constructor and resource fields, respectively. When attached to nodes, receiver and provider ports correspond to receiver and provider fields, respectively. However, when attached to boxes, this is reversed: receivers correspond to providers and providers correspond to receivers. Constructor fields are bijectively associated with interfaces that reside at the same level ($R_{I}$, depicted by contact). Correspondences must also be consistent with respect to $R_I$, in that if a field is associated with an interface, ports corresponding to the field must only be connected to boxes (via $R_{CA}$) and nodes (via $R_{CU}$) that correspond to that interface. Finally, in each interface, there is a \emph{connectivity relation} ($R_C$) between fields, covered in Section~\ref{sec:validity}.

\begin{figure}[!t]
    \footnotesize
    \begin{minipage}[t]{0.2in}
    \vspace{0pt}
    \raggedright
    (i)
    \end{minipage}
    \begin{minipage}[t]{0.9in}
    \vspace{0em}
    \includegraphics[width=0.9in]{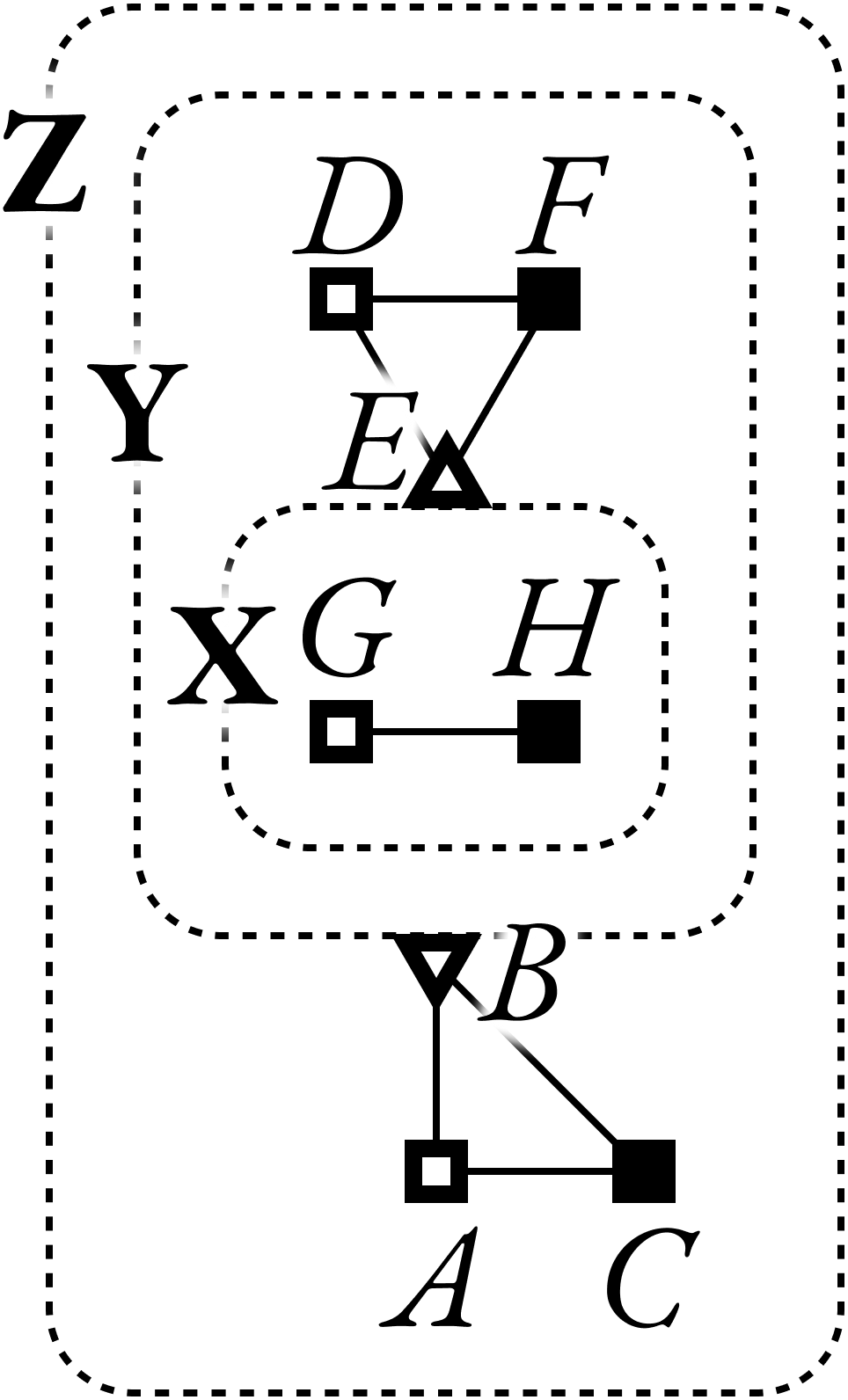}
    \end{minipage}
    \qquad
    \begin{minipage}[t]{0.2in}
    \vspace{0pt}
    \raggedright
    (ii)
    \end{minipage}
    \begin{minipage}[t]{1in}
    \vspace{-1.25em}
    \begin{align*}
    &\mathcal{I} = \{\mathbf{Z}, \mathbf{Y}, \mathbf{X}\} \\
    &\mathcal{F}_{C+} = \varnothing \\
    &\mathcal{F}_{C-} = \{B,E\} \\
    &\mathcal{F}_{R+} = \{C,F,H\} \\
    &\mathcal{F}_{R-} = \{A,D,G\} \\
    &R_I = \big\{(B, \mathbf{Y}), (E, \mathbf{X})\big\}
    \end{align*}
  \end{minipage}
  \begin{minipage}[t]{3in}
    \vspace{-1.25em}
    \begin{align*}
        &R_R = \Big\{
        \big(\mathbf{Z}, \{\mathbf{Y},A,B,C\}\big),
        \\ & \qquad
        \big(\mathbf{Y}, \{\mathbf{X},D,E,F\}\big), 
        \\ & \qquad
        \big(\mathbf{X}, \{G,H\}\big)
        \Big\} \\
        &R_C = \big\{
        (A,B), (A,C), (B,C),
        \\ & \qquad
        (D,E), (D,F), (E,F),
        (G,H)
        )\big\}
    \end{align*}
    \end{minipage}
\vspace{1em}
\\
    \begin{minipage}[t]{0.2in}
    \vspace{0pt}
    \raggedright
    (iii)
    \end{minipage}
    \begin{minipage}[t]{0.9in}
    \vspace{0pt}
    \includegraphics[width=0.9in]{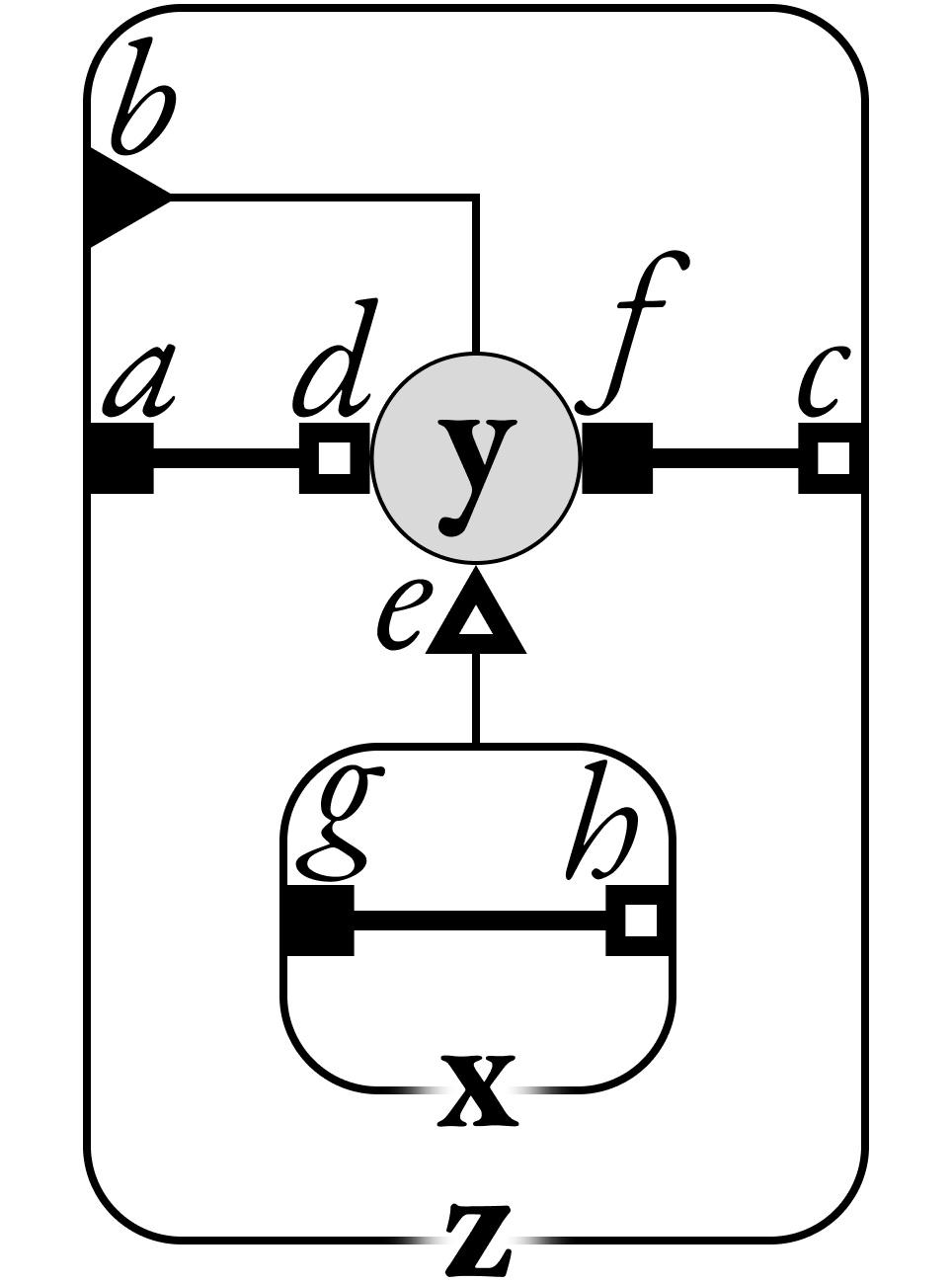}
    \end{minipage}
    \qquad
    \begin{minipage}[t]{0.15in}
    \vspace{0pt}
    \raggedright
    (iv)
    \end{minipage}
  \begin{minipage}[t]{1in}
    \vspace{-1.25em}
    \begin{align*}
&\mathcal{B} = \{\mathbf{z}, \mathbf{x}\} \\
&\mathcal{N} = \{\mathbf{y}\} \\
&\mathcal{P}_{C+} = \{b\} \\
&\mathcal{P}_{C-} = \{e\} \\
&\mathcal{P}_{R+} = \{a,f,g\} \\
&\mathcal{P}_{R-} = \{c,d,h\}
\end{align*}
\end{minipage}
  \begin{minipage}[t]{1in}
    \vspace{-1.25em}
    \begin{align*}
&\mathcal{D} = \varnothing \\
&\mathcal{T} = \varnothing \\
&R_{DI} = \varnothing \\
&R_{DC} = \varnothing \\
&R_{CA} = \big\{(\mathbf{x},e)\big\} \\
&R_{CU} = \big\{(\mathbf{y},b)\big\}
\end{align*}
\end{minipage}
\begin{minipage}[t]{2in}
    \vspace{-1.25em}
\begin{align*}
&R_{WC} = \varnothing \\
&R_{WR} = \big\{(a,d), (f,c), (g,h)\big\} \\
&R_R = \Big\{
\big(\mathbf{z}, \{\mathbf{y},\mathbf{x}, a,b,c,d,e,f\}\big),
\big(\mathbf{x}, \{g,h\}\big)
\Big\} \\
&R_A = \Big\{
\big(\mathbf{z}, \{a,b,c\}\big),
\big(\mathbf{y},\{d,e,f\}\big),
\big(\mathbf{x}, \{g,h\}\big)
\Big\}
\end{align*}
\end{minipage}
\vspace{1em} \\
\begin{minipage}[t]{1.45in}
\ 
\end{minipage}
\begin{minipage}[t]{0.2in}
\vspace{0pt}
\raggedright
(v)
\end{minipage}
\begin{minipage}[t]{2in}
    \vspace{-1.25em}
\begin{align*}
&C = \big\{(\mathbf{z},\mathbf{Z}),
(\mathbf{y}, \mathbf{Y}),(\mathbf{x}, \mathbf{X}), (a, A), (b, B), (c, C), (d, D), (e, E), (f, F), (g, G), (h, H)\big\}
\end{align*}
\end{minipage}

    \caption{\small An annotated illustration (i) and formalization (ii) of the type from Figure~\ref{fig:translation-table}~F. An annotated illustration (iii) and formalization (iv) of the term from Figure~\ref{fig:translation-table}~F. The correspondence between them (v), implicitly depicted via field and port positioning. $\{(a, \{b,c\})\}$ is shorthand for $\{(a,b), (a,c)\}$. See Figure~\ref{fig:typetermsoup} for the descriptions of all components of the formalism.}
    \label{fig:formalexg}
\end{figure}

\paragraph{Example.} In the illustrations, the depiction of correspondences is somewhat implicit, via the positioning of ports (either left, right, top, bottom, top-left, top-right, bottom-left, or bottom-right).
In Figure~\ref{fig:formalexg}, the correspondence $(a, A)$ is depicted by placing both the port and the field on the left of their containers, whereas $(b, B)$ is on the top left and $(c, C)$ is on the right. Since $\mathbf{Y}$ is associated with $B$ and $\mathbf{y}$ is constructed by $b$, the definition of correspondence relation forces the correspondence $(\mathbf{y}, \mathbf{Y})$. The placement of $d$ on the left of the node and $D$ on the left of the interface depicts the correspondence $(d,D)$, and similarly for $(e,E)$ on the bottom and $(f,F)$ on the right. The correspondence $(e,E)$ forces $(\mathbf{x}, \mathbf{X})$, and then the left and right positionings depict $(g,G)$ and $(h,H)$.

\paragraph{Types internal to terms.}
Recall the problem of associating ports between the body and occurrences of a let-binding:
the solution is to give each let-binding a type, and then, for the body and each occurrence of the let-binding, give a correspondence with the type.
In order to do so, terms themselves must contain types, which is accomplished via an \emph{internal type-fragment graph} ($\mathcal{T}$, not depicted), containing the unions of the vertices and edges of zero or more types. In particular, its residence relation ($R_{R}(\mathcal{T})$, not depicted), may be a forest rather than a tree. The interfaces in $\mathcal{I}(\mathcal{T})$ that are roots of this forest are in bijection ($R_{DI}$) with the let-bindings. Finally, the \emph{internal correspondence} ($R_{DC}$, depicted via relative port positioning) is a correspondence relating the body and occurrences of each let-binding with its associated interface. Now each port on an occurrence node is associated with a port on the body box, since they correspond to a shared field in the internal types.

\paragraph{Types external to terms.}
While the components deriving from let-bindings participate in the internal correspondence, those deriving from the root box of the term do not. Given a type $T$ and a term $t$, $C$ is an \emph{external correspondence between $t$ and $T$} if $C$ is a correspondence, if $C$ relates the root of $t$ to the root of $T$, and if $C$ is disjoint from $R_{DC}$.
When a term is paired with an external correspondence, every box, node, and port (except ports directly attached to let-bindings) corresponds to exactly one interface or field.

\paragraph{Canonicity of terms.} A term $t$ at a type $T$ in a functional language translates to, not just an \thelangname{} term, but the pair of an \thelangname{} term $\trans{t}_G$ and an external correspondence $\trans{t}_C$ between $\trans{t}_G$ and $\trans{T}$. Consider Figure~\ref{fig:translation-table}~B. In a traditional functional language, this type has two linear terms: \texttt{id := fun x, y => (x, y)} and \texttt{swap := fun x, y => (y, x)}. In \thelangname{}, there is only one term, which is shown, and is equal to $\trans{\mathtt{id}}_G = \trans{\mathtt{swap}}_G$. However, there are two distinct external correspondences between the term and the type, $\trans{\mathtt{id}}_C \neq \trans{\mathtt{swap}}_C$. When recursively translating terms, the inner correspondences cancel out, so we have both $\trans{\mathtt{id\ a\ b}}_G = \trans{\mathtt{swap\ b\ a}}_G$ and $\trans{\mathtt{id\ a\ b}}_C = \trans{\mathtt{swap\ b\ a}}_C$. This quotients out internal argument ordering while allowing $\mathtt{id}$ and $\mathtt{swap}$ to be differentiated externally.

\subsection{Connectivity Relation and Well-Formedness of Terms}
\label{sec:validity}
As mentioned previously, types have a connectivity relation, $R_C$. This is a symmetric, irreflexive relation among the fields residing in each interface. 
Intuitively, this relation indicates the allowed connections between ports on the ``inside'' of a node, and thus what might be wired together after substituting for the node.
This condition has several benefits: it rules out self-referential let-bindings that could lead to unbounded recursion; it allows us to faithfully represent traditional functions, where the output of a function may not be fed back as an input; and without it, our representations of data structures would not work correctly (see Figure~\ref{fig:bt-values}~(iv) and Figure~\ref{fig:abt-values}~(iii)).

\begin{figure}[!b]
    \footnotesize
    \centerline{\includegraphics[width=6.3in]{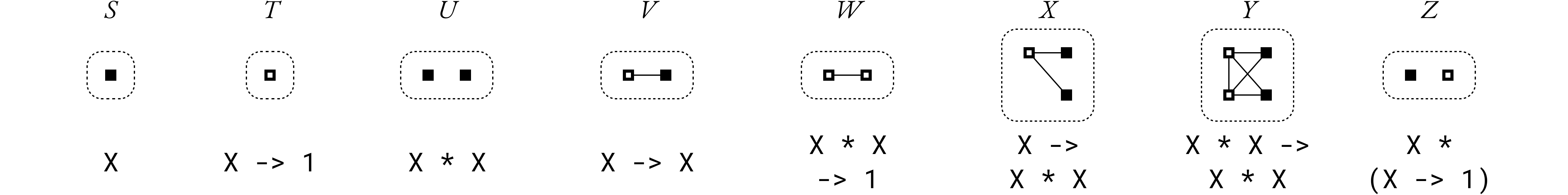}}
    \vspace{0.5em}
    \centerline{
        \hspace{0.9em}
        \makecell{$T^\perp$}
        \makecell{$S^\perp$}
        \hspace{0.05em}
        \makecell{$S \sqcup S$ \\ $W^\perp$}
        \hspace{0.09em}
        \makecell{$T \bowtie S$ \\ $Z^\perp$}
        \hspace{0.09em}
        \makecell{$T \bowtie T$ \\ $U^\perp$}
        \makecell{$T \bowtie U$}
        \makecell{$W \bowtie U$ \\ $T \bowtie X$}
        \makecell{$S \sqcup T$ \\ $V^\perp$ }
        \hspace{1.2em} \\
    }
    \caption{\small In each column, from top to bottom: a name; a type; a functional analogue; and one or two ways of forming the type from the other types with $\sqcup$, $\bowtie$, and $(\cdot)^\perp$.}
    \label{fig:typeformers}
\end{figure}

Roughly, the connectivity relation differentiates functions and products (more precisely $\otimes$ and $\parr$ in linear logic~\cite{linearlogic}, where linear functions are $A \multimap B \coloneqq A^\perp \parr B$). While a function may use its input to produce its output (its input and output ports may be connected), a product must produce its left and right sides separately (its left and right ports may not be connected). For a type $T$, define its \emph{dual}, $T^\perp$, to have the same fields but with, at the top-level, receivers and providers flipped and the complimentary connectivity relation. For types $T$ and $S$, define $T \sqcup S$ to have the disjoint union of the fields of $T$ and $S$ and the disjoint union of their connectivity relations. Define $T \bowtie S$ to be $T \sqcup S$, but adding connectivity edges between the top-level fields of $T$ and the top-level fields of $S$. When translating ordinary functional types to \thelangname{}, $\trans{A \times B}$ becomes $\trans{A} \sqcup \trans{B}$, and $\trans{A \to B}$ becomes $\trans{A}^{\perp} \bowtie \trans{B}$. See Figure~\ref{fig:typeformers} for examples.

\begin{figure}[!t]
    \footnotesize
    \centering
    \centerline{\includegraphics[width=6.3in]{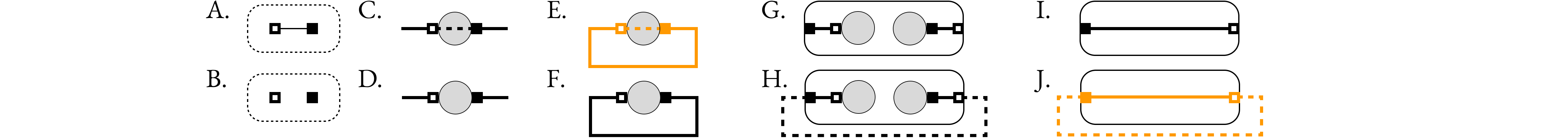}}
    \caption{\small Types (A, B) and fragments of terms (C, D, E, F, G, H, I, J). The interface in A corresponds to the nodes in C and E and to the boxes in G and I. The interface in B corresponds to the nodes in D and F and to the boxes in H and J.
    The dashed lines are not part of the term, but reflect $R_C$ between the fields of the type, shown between the corresponding ports. Note that because dual types are used for boxes, the dashed lines in G, H, I, and J are the compliment of $R_C$. E and J are ill-formed terms because of the cycles highlighted in orange.
    }
    \label{fig:acyclicity}
\end{figure}

For a term to be \emph{well-formed}, the combination of its wiring relation and the connectivity relation from its types must be acyclic, with certain exceptions (see Definition~\ref{defn:wellformed}). Figure \ref{fig:acyclicity} shows well- and ill-formed fragments of terms. There are four fragments, considered with two different types. The type for the top row is analogous to \texttt{X -> X}. C is valid, being analogous to an ordinary function call; E is invalid, analogous to a call where the output is fed back as the input; G is valid, analogous to a function that discards its argument by passing it to another function and then returning the result of an independent function call; I is valid, analogous to a function that returns its input as its output. The type for the bottom row does not have a perfect analogue in functional programming, but corresponds roughly to \texttt{(X -> 1) * X}: a pair of a continuation accepting an \texttt{X} and a value of type \texttt{X}. D is valid, analogous to using the continuation and the value  separately; F is valid, analogous to passing the value to the continuation; H is valid, analogous to constructing the continuation and value with separate function calls; J is invalid, analogous to returning the argument eventually passed to the continuation as the right side of the pair. Both E and J have prohibited cycles.

\begin{figure}[!b]
    \footnotesize
    \includegraphics[width=6.3in]{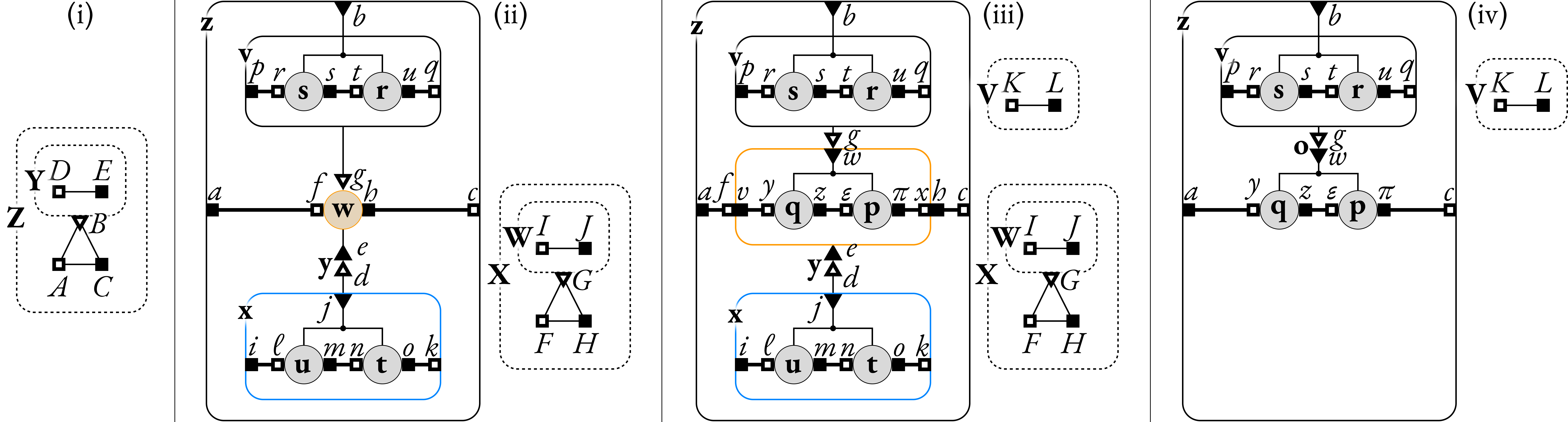}
    \vspace{0em}\\
    \begin{minipage}[t]{1in}
    \vspace{-2em}
    \begin{align*}
        &\mathcal{I}(\mathcal{T}) = \{\mathbf{X}, \mathbf{W}\} \\
        &\mathcal{F}(\mathcal{T}) = \{F, G, H, I, J\} \\
        &\mathcal{D} = \{\mathbf{y}\}  \\
        &R_A = \big\{(e, \mathbf{y}), (d, \mathbf{y}), ...\big\} \\
        &R_{DI} = \big\{(\mathbf{y}, \mathbf{X})\big\} \\
        &R_{WC} = \big\{(b, g)\big\}
    \end{align*}
    \end{minipage}
    \hfill
    \begin{minipage}[t]{1in}
    \vspace{-2em}
    \begin{align*}
        &R_{CU} = \big\{(\mathbf{s},b), (\mathbf{r},b), ...\big\} \\
        &R_{DC} = \big\{
            (\mathbf{w}, \mathbf{X}),
            (g, G),
            \\ & \qquad
            (\mathbf{v}, \mathbf{W}),
            (\mathbf{x}, \mathbf{X}),
            (j, G),
            \\ & \qquad
            (\mathbf{u}, \mathbf{W}),
            (\mathbf{t}, \mathbf{W}),
            ...
        \big\} \\
        &C = \big\{
            (\mathbf{z}, \mathbf{Z}),
            (b, B),
            \\ & \qquad
            (\mathbf{s}, \mathbf{Y}),
            (\mathbf{r}, \mathbf{Y}),
            ...
        \big\}
    \end{align*}
    \end{minipage}
    \hfill
    \begin{minipage}{0.3in}
    \raggedleft
    (v)
    \end{minipage}
    \hspace{-0.21em}
    \vline
    \hspace{0.21em}
    \begin{minipage}{0.3in}
    \raggedright
    (vi)
    \end{minipage}
    \hfill
    \begin{minipage}[t]{1in}
    \vspace{-2em}
    \begin{align*}
        &\mathcal{I}(\mathcal{T}) = \{\mathbf{V}\} \\
        &\mathcal{F}(\mathcal{T}) = \{K, L\} \\
        &\mathcal{D} = \{\mathbf{o}\} \\
        &R_A = \big\{(g, \mathbf{o}), (w, \mathbf{o}), ...\big\} \\
        &R_{DI} = \big\{(\mathbf{o}, \mathbf{V})\big\} \\
        &R_{WC} = \big\{(b, g)\big\}
    \end{align*}
    \end{minipage}
    \hfill
    \begin{minipage}[t]{1in}
    \vspace{-2em}
    \begin{align*}
        &R_{CU} = \big\{(\mathbf{s},b), (\mathbf{r},b), ...\big\} \\
        &R_{DC} = \big\{
            (\mathbf{v}, \mathbf{V}),
            (\mathbf{q}, \mathbf{V}),
            \\ & \qquad
            (\mathbf{p}, \mathbf{V}),
            ...
        \big\} \\
        &C = \{
            (\mathbf{z}, \mathbf{Z}),
            (b, B), 
            \\ & \qquad
            (\mathbf{s}, \mathbf{Y}),
            (\mathbf{r}, \mathbf{Y}),
        \ldots\}
    \end{align*}
    \end{minipage}

    \caption{\small An annotated step of the operational semantics at type (i), from term (ii) to term (iv) (shown previously in Figure~\ref{fig:operational-semantics-higher}). An illustration of an intermediate step, which is not a term (iii). The partial formalizations of the before term (v) and the after term (vi), and their correspondences $C$ to the type in (i). The omitted pieces of the formalizations are analogous to those in Figure~\ref{fig:formalexg}. The internal types $\mathcal{T}$, usually not depicted, are shown here. Term (ii) contains an internal interface $\mathbf{X}$, which is the type of the let-binding $\mathbf{y}$. Stepping at $\mathbf{y}$ substitutes the contents of $\mathbf{x}$ (the body of $\mathbf{y}$) for $\mathbf{w}$ (the occurrence of $\mathbf{y}$) and copies $\mathbf{W}$ to $\mathbf{V}$ (shown in (iii)). Finally, the pairs of resource ports $f$, $v$ and $x$, $h$ are replaced with wire, while $g$, $w$, and $\mathbf{V}$ are attached to a fresh let-binding, $\mathbf{o}$, yielding term (iv).}
    \label{fig:formalopsemformal}
\end{figure}

\subsection{Formalism}
\label{sec:formalism}

We now provide a precise formulation of \thelangname{}. We suggest referring to Figures~\ref{fig:formalexg} and ~\ref{fig:formalopsemformal} to ground definitions as they are introduced.

\begin{figure}[!b]
    \footnotesize
    \centering
    \begin{tabular}{c c p{0.54\linewidth}}
    \toprule
    \multicolumn{2}{l}{\textbf{Component} \hspace{0.46in} \textbf{Kind}} & \textbf{Description}\\
    \midrule
    $\mathcal{I}$ &
    set &
    \emph{interfaces}, depicted as dotted, rounded rectangles \\ \addlinespace[0.5em]
    $\mathcal{F}$ &
    set &
    \emph{fields}, divided into constructor ($\mathcal{F}_C$, triangle) or resource ($\mathcal{F}_R$, square) and provided ($\mathcal{F}_+$, solid) or received ($\mathcal{F}_-$, hollow) \\ \addlinespace[0.5em]
    $R_{R}$ &
    $(\mathcal{I} \rmoz \mathcal{I}) \cartprod (\mathcal{F} \rmo \mathcal{I})$, acyclic &
    \emph{residence}, depicted by containment in interfaces \\ \addlinespace[0.5em]
    $R_{I}$ &
    $\bigcartprod\limits_{i \in \mathcal{I}} \mathcal{F}_C^i \roo \mathcal{I}^i$ &
    \emph{constructor-interface} association, depicted by contact between fields and interfaces \\ \addlinespace[0.5em]
    $R_{C}$ &
    $\bigcartprod\limits_{i \in \mathcal{I}} \operatorname{cograph\,on} \mathcal{F}^i$ &
    \emph{connectivity}, depicted with solid lines; \emph{cograph} is Definition~\ref{defn:cograph} \\ 
    \midrule
    $\mathcal{B}$ &
    set &
    \emph{boxes}, depicted as solid, rounded rectangles \\ \addlinespace[0.5em]
    $\mathcal{N}$ &
    set &
    \emph{nodes}, depicted as gray circles or rectangles \\ \addlinespace[0.5em]
    $\mathcal{P}$ &
    set &
    \emph{ports}, divided into constructor ($\mathcal{P}_C$, triangle) or resource ($\mathcal{P}_R$, square) and provided ($\mathcal{P}_+$, solid) or received ($\mathcal{P}_-$, hollow) \\ \addlinespace[0.5em]
    $\mathcal{D}$ &
    set &
    \emph{let-bindings}, depicted as contact between constructor ports \\ \addlinespace[0.5em]
    $\mathcal{T}$ &
    type-fragment graph &
    \emph{internal type-fragment graph}, not depicted \\ \addlinespace[0.5em]
    $R_{R}$ &
    $(\mathcal{B} \rmoz \mathcal{B}) \cartprod (\mathcal{N} \cup \mathcal{P} \cup \mathcal{D} \rmo \mathcal{B})$, acyclic &
    \emph{residence}, depicted by containment in boxes \\ \addlinespace[0.5em]
    $R_A$ &
    $\bigcartprod\limits_{b \in \mathcal{B}} \mathcal{P}^b \rmo (\mathcal{N}^b \cup \mathcal{D}^b \cup \{b\})$ &
    \emph{attachment}, depicted by contact between ports and nodes/boxes \\ \addlinespace[0.5em]
    $R_{WR}$ &
    $\bigcartprod\limits_{b \in \mathcal{B}} \mathcal{P}_{R+}^b \roo \mathcal{P}_{R-}^b$ &
    \emph{resource wiring}, depicted with thick lines \\ \addlinespace[0.5em]
    $R_{WC}$ &
    $\bigcartprod\limits_{b \in \mathcal{B}} \mathcal{P}_{C+}^b \rmm \mathcal{P}_{C-}^b$ &
    \emph{constructor wiring}, not depicted \\ \addlinespace[0.5em]
    $R_{CA}$ &
    $\bigcartprod\limits_{b \in \mathcal{B}}  \mathcal{B}^b \roo \mathcal{P}_{C-}^b$ &
    \emph{constructor argument}, depicted with thin lines \\ \addlinespace[0.5em]
    $R_{CU}$ &
    $\bigcartprod\limits_{b \in \mathcal{B}} \mathcal{N}^b \rmo \bigcup\limits_{b \sqsubseteq b'} \mathcal{P}_{C+}^{b'}$, wire-safe &
    \emph{constructor usage}, depicted with thin, possibly branching lines; \emph{wire-safe} is Definition~\ref{defn:wiresafe}  \\ \addlinespace[0.5em]
    $R_{DI}$ &
    $\mathcal{D} \roo \operatorname{roots\ of} \mathcal{I}(\mathcal{T})$ &
    \emph{let-binding typing}, not depicted \\ \addlinespace[0.5em]
    $R_{DC}$ &
    $\bigcartprod\limits_{d \in \mathcal{D}} \operatorname{correspondence\ for} d$ &
    \emph{let-binding correspondence}, depicted by the positioning of ports on boxes and nodes; \emph{correspondence for $d$} is Definition~\ref{defn:letcorrespondence} \\ 
    \bottomrule
    \end{tabular}
    \caption{\small Components of a type-fragment graph (top) and a term-fragment graph (bottom). $\mathcal{R} \roo \mathcal{S}$ is a one-to-one relation,
    $\mathcal{R} \rmo \mathcal{S}$ is many-to-one, $\mathcal{R} \rmoz \mathcal{S}$ is many-to-one-or-zero, and $\mathcal{R} \rmm \mathcal{S}$ is many-to-many. $\mathbf{S} \cartprod \mathbf{R} = \{S \cup R : S \in \mathbf{S}, R \in \mathbf{R}\}$. $\mathcal{S}^i$ and $\mathcal{S}^b$ denote the subsets of $\mathcal{S}$ residing directly in $i$ and $b$. $c \sqsubseteq i$ and $c \sqsubseteq b$ mean that $c$ is below $i$ and $b$ in the residence forest. $\mathcal{F}_{C-} = \mathcal{F}_C \cap \mathcal{F}_-$, and likewise for $\mathcal{F}_{C+}$, $\mathcal{F}_{R-}$, $\mathcal{F}_{R+}$, $\mathcal{P}_{C-}$, $\mathcal{P}_{C+}$, $\mathcal{P}_{R-}$, and $\mathcal{P}_{R+}$.}
    \label{fig:typetermsoup}
\end{figure}

\begin{definition}[fragment graphs]
We define \emph{type-fragment graphs} and \emph{term-fragment graphs} in Figure~\ref{fig:typetermsoup}. Each consists of several sets of vertices and several edge relations with conditions.
\end{definition}

\begin{definition}[types and terms]
A type-fragment graph is a \emph{type} if $R_R$ has a single root interface. A term-fragment graph is a \emph{term} if $R_R$ has a single root box.
\end{definition}

\begin{definition}[cographs]
\label{defn:cograph}
The set of \emph{cographs on $\mathcal{V}$} is the smallest set of symmetric, irreflexive graphs on vertices $\mathcal{V}$ that contains the singleton graphs and is closed under complement and disjoint 
union. Intuitively, this is the set of formulas on atoms $\mathcal{V}$ that can be formed with conjunction and disjunction, quotienting out associativity and commutativity.
\end{definition}

\begin{definition}[wire-safety]
\label{defn:wiresafe}
Assume a constructor usage relation $R_{CU}$, a constructor wiring relation $R_{WC}$, and a pair $(n, c) \in R_{CU}$. Let $c$ reside in $b_c$ and $n$ reside in $b_n$, where $b_n \sqsubseteq b_c$. If $b_n \neq b_c$, let $b_n'$ be the box residing directly in $b_c$ such that $b_n \sqsubseteq b_n' \sqsubset b_c$, and let $c'$ be the constructor receiver port (in $b_c$) associated with $b_n'$. $R_{CU}$ is \emph{wire-safe} for $R_{WC}$ if, for all $(n, c) \in R_{CU}$, either $b_n = b_c$ or $(c, c') \in R_{WC}$.
\end{definition}

\begin{definition}[correspondences]
\label{defn:correspondence}
Given a type $T$ and a term $t$, a relation
$C \in (\mathcal{B} \rmoz \mathcal{I}) \cartprod (\mathcal{N} \rmoz \mathcal{I}) \cartprod (\mathcal{P} \rmoz \mathcal{F})$
is a
a \emph{correspondence} if the following hold:
(1) If $b \in \mathcal{B}$ (or $n \in \mathcal{N}$) corresponds to $\iota \in \mathcal{I}$, then the correspondence relation is bijective between the ports attached to $b$ (or $n$) and the fields of $\iota$.
(2) If $p \in \mathcal{P}_{C}$ corresponds to $f \in \mathcal{F}_{C}$, then the box associated with $p$ corresponds to the interface associated with $f$.
(3) Constructor and resource ports are associated to constructor and resource fields, respectively.
(4) If $p$ corresponds to $f$, their receiver and provider kinds are the same if $p$ is attached to a node and opposite if $p$ is attached to a box.
\end{definition}

\begin{definition}[let-binding correspondences]
\label{defn:letcorrespondence}
A correspondence $C$ is a \emph{correspondence for $d \in \mathcal{D}$} if the box of $d$ (via $R_{A}$ and $R_{CA}$) and nodes of $d$ (via $R_{A}$ and $R_{CU}$) correspond to the interface of $d$ (via $R_{DI}$). Distinct let-binding correspondences must cover disjoint sets of term components.
\end{definition}

\begin{definition}[external correspondences]
Given a type $T$ an a term $t$, we say that a correspondence relation $C$ is an \emph{external correspondence} between $t$ and $T$ if $C$ relates the root box of $t$ with the root interface of $T$ and $C$ is disjoint from $R_{DC}$.
\end{definition}

\noindent \textbf{Remark.} Given a type $T$, a term $t$, and an external correspondence $C$ between $t$ and $T$. Let $C^* \coloneqq C \cup R_{DC}$. Every $n \in \mathcal{N}$ and $b \in \mathcal{B}$ occurs exactly once in $C^*$. For every $p \in \mathcal{P}$, either it is attached to some $d \in \mathcal{D}$ and does not occur in $C^*$, or it occurs exactly once in $C^*$.

\begin{definition}[term equality]
Given a type $T$, terms $t_1$ and $t_2$, and correspondences $C_1$ between $t_1$ and $T$ and $C_2$ between $t_2$ and $T$, we say that $(t_1, C_1)$ is \emph{$T$-equal} to $(t_2, C_2)$ if, fixing a concrete labeling of vertices to yield $\widehat{T}$, $\widehat{t_1}$,  $\widehat{t_2}$, $\widehat{C_1}$, and $\widehat{C_2}$, there exists some relabeling $h$ of the vertices in $\widehat{t_2}$ such that $(\widehat{t_1}, \widehat{C_1}) = \big(h(\widehat{t_2}), h(\widehat{C_2})\big)$.
\end{definition}

\begin{definition}[substitution]
Assume a type $T$, term $t$, and correspondence $C$ between $t$ and $T$.
Given $b \in \mathcal{B}$ and $n \in \mathcal{N}$, where $b$ resides in some $b_0 \in \mathcal{B}$ and $n$ is at or below $b_0$ in the residence forest, and given $\iota \in \mathcal{I}(\mathcal{T})$ and correspondences $C_b \subseteq R_{DC}$ between $b$ and $\iota$ and $C_n \subseteq R_{DC}$ between $n$ and $\iota$,
we define the \emph{substitution of $(b, C_b)$ for $(n, C_n)$ at $\iota$} to be the result if we:
(1) Delete $n$ (from $\mathcal{N}$ and all relations).
(2) For each component residing in $b$, make a fresh copy residing in the box that contained $n$, also making appropriate copies in $R_{DC}$ and $C$.
(3) For each port $p_b$ attached to $b$, let $p_b'$ be the fresh copy of $p_b$, and let $C_{bp}' \subseteq R_{DC}$ be the portion relevant to $p_b'$.
Note that, for each $f$ residing in $\iota$, we now have a $p_b'$ that is fresh and a $p_n$ that used to be attached to $n$, and that they are a received/provided pair. Let $C_{np} \subseteq C_n$ be the part relevant to $p_n$.
(4) For each resource field $f$, $p_b'$ was wired to some $p_b''$ and $p_n$ was wired to some $p_n'$. Erase $f$, $p_b'$, and $p_n$ and add $(p_b'',p_n')$ to the wiring relation.
(5) For each constructor field $f$, create a new let-binding $d$, and attach $p_b'$ and $p_n$ to it. For the $\iota_f \in \mathcal{I}(\mathcal{T})$ associated with $f$, make a fresh copy of its subtree in $\mathcal{I}(\mathcal{T})$ and let the root of the copy be $\iota_f'$. Change $C_{bp}'$ and $C_{np}$ to refer to $\iota_f'$. Add $(d, \iota_f')$ to $R_{DI}$.
\end{definition}

\begin{definition}[inlining of let-bindings]
Assume a type $T$, term $t$, and correspondence $C$ between $t$ and $T$. Given a let-binding $d \in \mathcal{D}$, let the associated interface be $\iota$, the attached constructor receiver port be $p_-$, the attached constructor provider port be $p_+$, the argument to $p_-$ be the box $b$, the set of nodes produced by $p_+$ be $N$, the subset of $R_{DC}$ relevant to $b$ be $C_b$, and for each $n \in N$, the subset of $R_{DC}$ relevant to $n$ be $C_n$. Define the \emph{inlining of $d$} to be the result if we:
(1) For each $n \in N$, substitute $(b, C_b)$ for $(n, C_n)$ at $\iota$.
(2) Delete $d$, $\iota$, $p_-$, $p_+$, and $b$.
\end{definition}

\begin{definition}[reduction]
Given a type $T$, terms $t_1$ and $t_2$, and correspondences $C_1$ between $t_1$ and $T$ and $C_2$ between $t_2$ and $T$, we say that $(t_1, C_1)$ \emph{$T$-reduces} to $(t_2, C_2)$ if there exists a $d \in \mathcal{D}(t_1)$ such that the result of inlining $d$ in $(t_1, C_1)$ is $T$-equal to $(t_2, C_2)$.
\end{definition}

\begin{definition}[descent of components]
A \emph{component} is a box, node, port, or let-binding. A component $c_1$ \emph{is a child of} $c_2$ if $c_1$ is attached (related in $R_A$) to $c_2$, if $c_1$ is the constructor argument (related in $R_{CA}$) of $c_2$, or if $c_1$ is a constructor usage (related in $R_{CU}$) of $c_2$. For the transitive closure of this relation, $c_1$ \emph{descends from} $c_2$. Note that the descent relation forms a forest, separate from the residence forest ($R_R$), and note that every component descends either from a let-binding or one of the roots of the residence forest.
\end{definition}

\begin{definition}[well-formedness]
\label{defn:wellformed}
Assume a type $T$, a term $t$, and an external correspondence $C$ between $t$ and $T$. Let $W$ be the relation on ports $R_{WR} \cup R_{WC}$. Define the relation $F$ on ports such that $(p_1, p_2) \in F$ if either: (1) $p_1$ and $p_2$ are attached to the same $d \in \mathcal{D}$; (2) let $c$ be the nearest common ancestor box or node of $p_1$ and $p_2$ in the descent relation, let $p_1',p_2'$ be the ancestors of $p_1,p_2$ (respectively) attached to $c$, and let $f_1',f_2'$ be the fields corresponding to $p_1',p_2'$ (respectively); if there is no such $c$, $(p_1, p_2) \notin F$; otherwise, if $c$ is a node, then $(p_1, p_2) \in F$ if $(f_1', f_2') \in R_C$;  if $c$ is a box, then $(p_1, p_2) \in F$ if $(f_1', f_2') \notin R_C$. Now, we say $(T,t,C)$ is \emph{well-formed} if for every cycle taking alternating edges in $F$ and $W$, there is some pair of ports $p_1,p_2$ in this cycle such that $(p_1,p_2) \in F$ but the edge $(p_1,p_2)$ is not part of the cycle. (This is directly inspired by the chorded-acyclic R\&B-cograph condition from \cite{handsome}, and extended to handle the nesting of interfaces.)
\end{definition}

\section{Expressing Data}
\label{sec:repr}
\subsection{Binary Trees}
\label{sec:repr:bt}
\begin{figure}[!b]
    \footnotesize
    \centering
    \centerline{\includegraphics[width=6.3in]{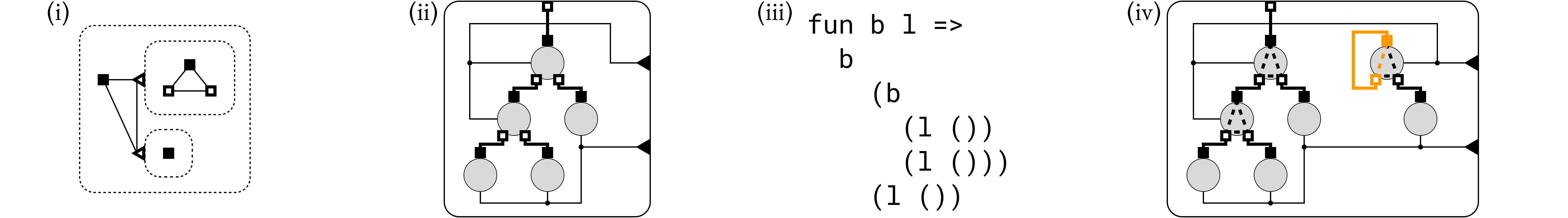}}
    \caption{\small The type of unlabeled binary trees (i). A term of that type (ii). The representation of that term as a Church-encoding in a generic functional language (iii). A term that is ill-formed as an unlabeled binary tree (iv). The dashed lines are not part of the term, but are $R_C$ between the fields of the type, shown between the corresponding ports. The illegal cycle is marked in orange.}
    \label{fig:bt-values}
\end{figure}

Now we will see how \thelangname{} represents data, using unlabeled binary trees as an example. In polymorphic lambda calculus, they are represented by the type $\forall X.  (X \to X \to X) \to (1 \to X) \to X$: the first argument, $X \to X \to X$, is the (arity-2) branch constructor, and the second argument, $1 \to X$, is the (arity-1) leaf constructor (taking unit, since the trees are unlabeled). Dropping polymorphism, this corresponds to the \thelangname{} type in Figure \ref{fig:bt-values}~(i). The top-right constructor field represents branch nodes, with resource ports for one parent (top), one left child (bottom-left), and one right child (bottom-right); the bottom-right constructor field represents leaf nodes, with a resource port for one parent (top). The remaining resource port (top-left) corresponds to the root of the tree.

Figure~\ref{fig:bt-values}~(ii) and~(iii) represent the same binary tree, with two branch nodes and three leaf nodes. The connectivity relation and well-formedness condition rule out terms like in Figure~\ref{fig:bt-values}~(iv). Linearity ensures that there is a single tree: additional trees would have no resource port to serve as their root, and such terms would be ruled out by the bijectivity of the resource wiring relation.

\subsection{Directed Multigraphs}
\label{sec:repr:dmg}

\begin{figure}[h]
    \footnotesize
    \centering
    \centerline{\includegraphics[width=6.3in]{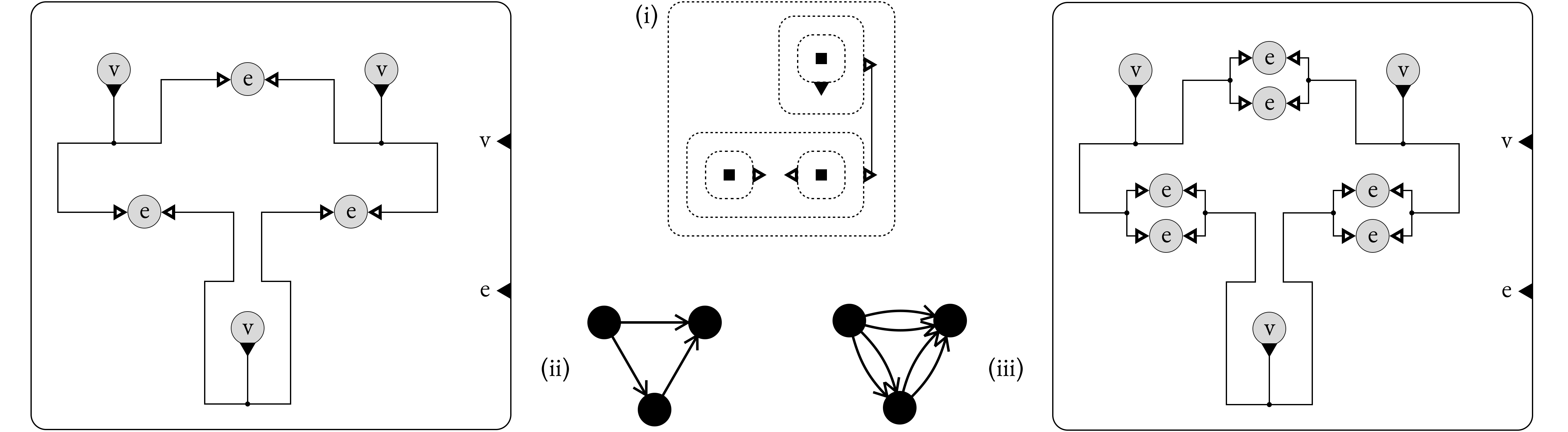}}
    \caption{\small The type of directed multigraphs (i). Two terms and the directed multigraphs they represent, (ii) and (iii). To improve readability, the constructor usage relation is depicted with the labels ``v'' and ``e'' rather than lines.}
    \label{fig:mg-values}
\end{figure}

As an example of a structure that is not a term algebra, and thus lacks a traditional Church-encoding~\cite{encodetermalgebras}, consider directed multigraphs.
They are specified in \thelangname{} by the type in Figure~\ref{fig:mg-values}~(i), with the top-right field corresponding to ``vertices'' and the bottom-right field corresponding to ``edges''. 
Figure~\ref{fig:mg-values}~(ii) shows a multigraph with three vertices and three edges and its representation as a term. Because vertices may be associated with any number of edges, vertex nodes have a constructor port rather than a resource port. Edges have two ports, for receiving constructors from their source and target vertices. Here, constructor provider ports are shown directly connected to constructor receiver ports, which is not formally allowed---we abuse notation for clarity, and mean that the receiver port is attached to a box which contains a single node constructed by the provider port. Figure~\ref{fig:mg-values}~(iii) shows a similar multigraph with three vertices, but with six edges. There is a related type for representing bags, which are essentially multigraphs without edges.

\subsection{Untyped Lambda Calculus}
\label{sec:repr:abt}
Closed terms in untyped lambda calculus also do not form a term algebra. They have three kinds of nodes---application, abstraction, and variable---but every variable node must somehow be associated with an abstraction above it.

\begin{figure}[t!]
    \footnotesize
    \centering
    \centerline{\includegraphics[width=6.3in]{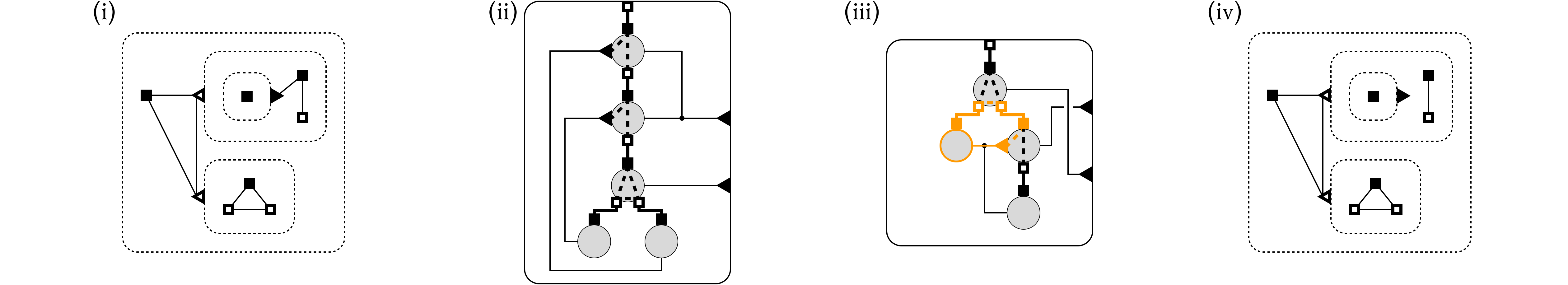}}
    \caption{\small The type of closed terms in untyped lambda calculus (i). A term representing ``$\lambda x. \lambda y.\,y\ x$'' (ii). A term corresponding to ``$x\ (\lambda x. x)$'' (iii). Another type, differing from (i) by a single connectivity edge (iv). The connectivity relation from (i) is overlaid with dashed lines on the corresponding ports in (ii) and (iii). The cycle marked in orange makes the external correspondence between (i) and (iii) ill-formed. There is a well-formed external correspondence between (iii) and (iv).}
    \label{fig:abt-values}
\end{figure}

Figure~\ref{fig:abt-values}~(i) shows their type in \thelangname{}, with the bottom-right constructor field representing application, having resource fields for a parent (top), a left child (bottom-left), and a right child (bottom-right), and the top-right constructor field representing abstraction, having resource fields for a parent (top) and a child (bottom), and a constructor port for ``variable nodes referring to this abstraction'' (left). Rather than starting with a single constructor for ``variable'' nodes, each time an abstraction node is constructed, a new ``variable'' constructor appears. The connectivity relation on the interface of abstractions prevents variables from occurring above their binder, as in Figure~\ref{fig:abt-values}~(iii): the edge between the ``variable'' constructor port and the ``parent'' port prohibits this, while the lack of edge between the ``variable'' constructor port and the ``child'' port allow variables to occur in the body of an abstraction. In contrast, Figure~\ref{fig:abt-values}~(iv) lacks the variable-parent edge and thus does admit this term.

This representation is closely related to parametric higher-order abstract syntax (\textsc{phoas})~\cite{bgb,phoas}, which leverages parametric polymorphism to represent variable binding. Computation over our representation, like \textsc{phoas}, respects the binding structure of terms, allowing the implementation of single-step beta-reduction and providing capture-avoiding substitution for free.

\section{Manipulating Data}
\label{sec:manip}

\begin{figure}[!b]
    \footnotesize
    \centering
    \includegraphics[width=6.3in]{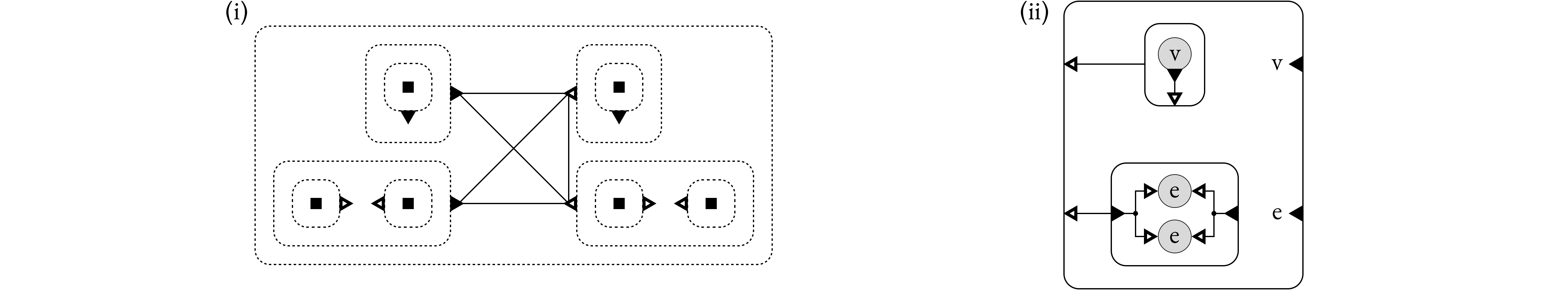}
    \caption{\small The type of functions from directed multigraphs to directed multigraphs (i). The term of that type that replaces each edge in the input with a pair of edges (ii).}
    \label{fig:mg-transformer}
\end{figure}

Now we show how to manipulate such data structures.
The lack of polymorphism in this simplified presentation forces a simple example,
since it is not clear how to write many functions over Church-encodings without instantiating the universal quantifier with complex types.
Though the full version of \thelangname{} can represent much richer functions, the following example should provide the right intuition.

\begin{figure}[!t]
    \footnotesize
    \centering
    \centerline{\includegraphics[width=6.3in]{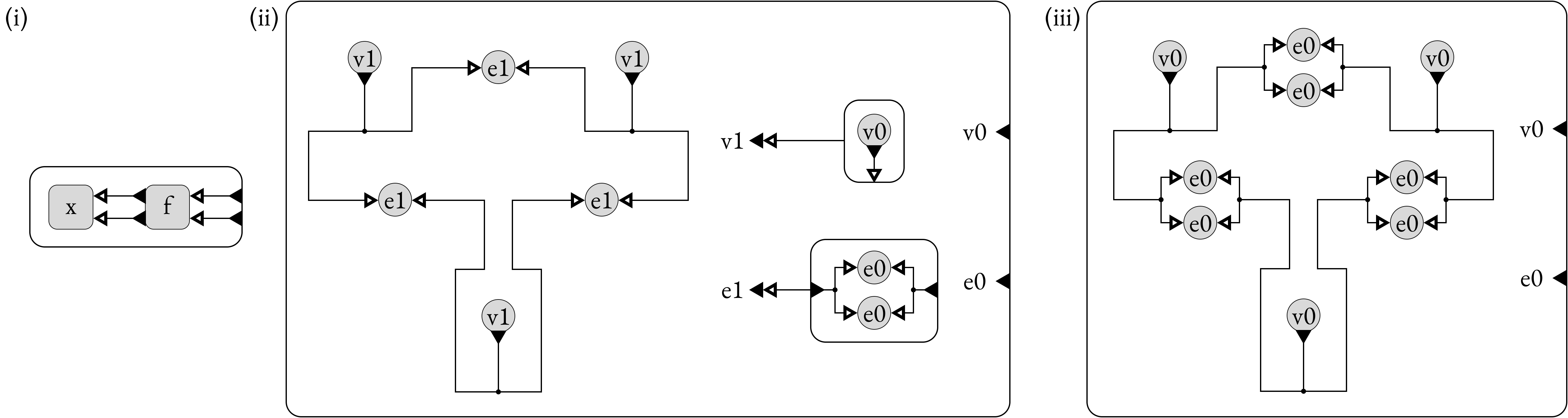}}
    \caption{\small The term that passes an instance of ``x'' to a call to ``f'' (i). The result of inlining in (i) the definitions of ``x'' and ``f'' (ii). The result of inlining in (ii) the let-bindings ``v1'' and ``e1'' (iii). ``x'' is a let-binding whose body is the directed multigraph in Figure~\ref{fig:mg-values}~(ii), and ``f'' is a let-binding whose body is the function in Figure~\ref{fig:mg-transformer}~(ii). (iii) is the term from Figure~\ref{fig:mg-values}~(iii).}
    \label{fig:mg-reduction}
\end{figure}

Recall the representation of directed multigraphs from Section~\ref{sec:repr:dmg}. The term in Figure~\ref{fig:mg-transformer}~(ii) behaves like a function that takes a directed multigraph as input and doubles each edge. Figure~\ref{fig:mg-reduction} depicts the evaluation of this function on the directed multigraph with three vertices and three edges from Figure~\ref{fig:mg-values}~(ii).

We emphasize that, in most programming languages, manipulating a graph involves manipulating a structure with labeled nodes---be it an adjacency matrix or a list of pairs of node indices---which makes it possible to write functions that are dependent on the labeling, and thus not truly functions over graphs. \thelangname{} cannot do that: it merely replaces each node of a structure (in this case both vertices and edges are nodes) with some pattern, as with Church-encodings; in this case, each vertex is replaced by a single vertex, and each edge by two edges of the same orientation.

Unfortunately, this is conservative: there are legitimate functions on graphs that we cannot express, including the function that returns the number of vertices as a Church-numeral. Specifically, \thelangname{} has two distinct types that resemble the natural numbers, which are roughly ``lists of units'' and ``bags of units''; we can write the function that counts the number of vertices as a bag of units, but we cannot write the function that converts a bag of units to a list of units, even though it would be sound. Though we could add a primitive function to accomplish this, we might wish to define it internally. Characterizing and enlarging the set of functions that can be represented is left for future work.

\section{Related Work}
\label{sec:relatedwork}

\paragraph{Linear logic.}
\thelangname{} is closely related to linear logic~\cite{linearlogic}.
Most presentations of linear logic use the rules of ``contraction'', ``weaking'', and ``dereliction'' for exponentials, but Andreoli's equivalent dyadic system~\cite{focusing} instead uses a rule called ``adsorption'', which is very reminiscent of our nodes and our constructor usage relation. An obvious difference is that our propositions are graphs, not trees, allowing us to quotient out certain type equivalences, like the ordering of products. The type equivalences that we quotient out are similar to the \emph{provable type isomorphisms} for intuitionistic type systems~\cite{typeisos}. This leads us to conjecture that \thelangname{} is polymorphic (second-order propositional) multiplicative exponential linear logic with the \textsc{mix} rule, but with these type isomorphisms quotiented out.

\relatedworkspacefudge
\paragraph{Proof nets and interaction nets.}
Our work is closely related to proof nets~\cite{linearlogic}, and, in particular, their extension, interaction nets~\cite{interactionnets}. There are two key differences between our work and interaction nets: (1) In interaction nets, each symbol (roughly our ``node'') has a \emph{principal} port, which is used in reduction; in our work, nodes do not have privileged ports, and reduction proceeds exclusively by substituting definitions (of types or terms) for their occurrences. (2) In interaction nets, the set of symbols and their associated ports must be fixed ahead of time; in our work, the symbol set is not fixed, with occurrences of symbols potentially adding fresh symbols to the set.

There is work on representing lambda calculus terms using interaction nets~\cite{geometryofimplementation,interactionnetimpl,lambdascope}. This work uses explicit ``duplication'' and ``erasure'' symbols, whereas exponentials (``constructors'' in our terminology) are a central piece of our formalism.
A key advantage of that work is improved reduction performance on some benchmarks, facilitated by the sharing of subterms~\cite{interactionnetimpl}. We hope to evaluate our system on their benchmark in future work.

There are versions of both proof nets~\cite{linearlogic} and interaction nets~\cite{pnin,inll} that represent exponentials with ``boxes'', and we expect these to be closely related to \thelangname{}, though our types are graphs rather than trees.

\relatedworkspacefudge
\paragraph{Functional programming.}
There are several key differences between \thelangname{} and more traditional functional programming languages like OCaml, Haskell, and Rust. (1) \thelangname{} does not have primitive inductive datatypes, instead using an analogue of Church-encodings. (2) \thelangname{} is both pure and strongly-normalizing and does not prescribe an evaluation order. (3) \thelangname{} is linear in the sense of Girard~\cite{linearlogic}, whereas Rust and Linear Haskell lack exponentials, Linear Haskell has separate non-linear types, and Rust is affine. (4) Most languages have functions implicitly return a single value, but boxes (``functions'') in \thelangname{} explicitly name their zero or more outputs, similar to out-parameters in C and similar languages. (5) \thelangname{} has a natural interpretation as graph substitution, even if a textual formalism were preferred for writing programs.

\relatedworkspacefudge
\paragraph{Polymorphic lambda calculus.}
There are three key differences between (polymorphic) \thelangname{} and polymorphic lambda calculus (System F): \thelangname{} is conjectured to be a canonical version of polymorphic multiplicative exponential linear logic (\textsc{pmell}) with the \textsc{mix} rule; \textsc{pmell} is the classical counterpart to polymorphic intuitionistic linear logic (\textsc{pill}); and \textsc{pill} is the linear counterpart to System F. The presentation here is not polymorphic, and so corresponds to intuitionistic linear logic and the simply-typed lambda calculus.

In terms of ability to express data types, we expect Ideograph and \textsc{pmell} to be the same. However, the canonicity of Ideograph means that e.g. for a directed multigraph $g$, where in \textsc{pmell} there is a different (fully-normalized) term for each labeling of the vertices and edges of $g$, in Ideograph there is a unique (fully normalized) term representing $g$. We are unsure of how linearity and classicality affect the ability to express data (i.e. the set of types and their fully-normalized terms). Linearity and classicality have established effects on the computational behavior of languages: linear calculi are often able to explicitly distinguish call-by-value and call-by-name~\cite{twotrans}, and classicality allows the expression of constructs like call/cc~\cite{calcc}. Data structures that form heterogeneous term algebras can be procedurally Church-encoded into System F types~\cite{encodetermalgebras}, but structures like directed multigraphs and lambda calculus terms are not term algebras.

\relatedworkspacefudge
\paragraph{Graph representations of programming languages.}
There is work representing existing programming languages, in particular lambda calculus, as graphs~\cite{hypergraph,categoricalclosureconversion,twoillustrations}. A key difference of our work is that we are not trying to represent existing programming languages for the purposes of, e.g. optimizing compilation. Rather we want to represent data structures (of which syntax trees happen to be one) and pure computations over them. As a result, we are not concerned with effects or evaluation order, with which much of this work contends.

\relatedworkspacefudge
\paragraph{Graph representations of types.}
There is work on representing formulas in multiplicative linear logic as undirected graphs~\cite{mllgraphs,handsome}. Our type system  is closely related to these when we do not use exponentials or second-order propositional quantifiers. (Note that we adopt the edge convention opposite of theirs for $\otimes$ and $\parr$.)

\relatedworkspacefudge
\paragraph{Representations of graphs.}
There is work representing graph structures in pure functional programming languages via recursive binders~\cite{structuredgraphs}. While our system is linearly typed and theirs is not, we expect them to be closely related and hope to pursue the connection in future work.

\relatedworkspacefudge
\paragraph{Graph programming languages.}
There are several graph programming languages, including GROOVE~\cite{groove}, GP-2~\cite{gp2}, and LMNtal~\cite{lmntal}.
All such systems we are aware of have a notion of a rewrite rule, which matches a subgraph and replaces it with some other subgraph. In contrast, our system has only one reduction rule, which is analogous to beta reduction. LMNtal lacks a type system, whereas types are a core part of \thelangname{}. HyperLMNtal~\cite{hyperlmntal}, which extends LMNtal with hyperedges, has been used to encode lambda calculus terms. In contrast to our use of constructors, they connect a binder to all of its variable occurrences via a single hyperedge. \thelangname{} is a variant of ``labeled port graphs''~\cite{portgraph}, a formalism where edges connect to nodes at ``ports'', which has been used to represent programs.

\relatedworkspacefudge
\paragraph{Parametric higher-order abstract syntax.}
There is work on encoding syntax with variable binding using functions in the meta-language. In particular, \cite{bgb} and \cite{phoas} leverage parametric polymorphism to encode exactly the closed terms in several lambda calculi and allow only structure-respecting operations. Their encodings, when translated into the types of \thelangname{}, correspond very closely to the type we presented in Section \ref{sec:repr:abt}. However, \thelangname{} can represent languages where variables must be used exactly once, and parametric higher-order abstract syntax cannot. Moreover, our goal is to represent structures beyond just syntax. 

\section{Future Work}
\label{sec:futurework}
There are several avenues for future work. We must first prove standard properties about \thelangname{}, including subject reduction and strong normalization. We would then like to prove that term-equality is graph isomorphism-complete, to characterize the structures that can be represented by the types, and to formally establish the connection to linear logic.
Finally, we plan to develop an implementation, which we expect to be fairly straightforward due to the simplicity of the operational semantics.

\section*{Acknowledgements}
We would like to thank Ian Mackie, Kazunori Ueda, and two anonymous reviewers for their valuable feedback, as well as Lawrence Dunn, Harrison Goldstein, Eleftherios Ioannidis, Nick Rioux, and Lucas Silver for reading early drafts. This work is funded in part by NSF Awards CCF-1910769 and CCF-1917852.

\nocite{*}
\bibliographystyle{eptcs}
\bibliography{refs}
\end{document}